\documentstyle[12pt]{article}
\begin{document}

%%%%%%%%%%%%%%%%%%%%%%%%%%%%%%%%%%%%%%%%%%%%%%%%%%%%%%%%%%%%%%%%%%%%%%%%%%%%%

\renewcommand{\thesection}{\arabic{section}}
\renewcommand{\theequation}{\arabic{equation}}
\renewcommand {\c}  {\'{c}}
\newcommand {\cc} {\v{c}}
\newcommand {\s}  {\v{s}}
\newcommand {\CC} {\v{C}}
\newcommand {\C}  {\'{C}}
\newcommand {\Z}  {\v{Z}}

\baselineskip=24pt

%%%%%%%%%%%%%%%%%%%%%%%%%%%%%%TITLE PAGE%%%%%%%%%%%%%%%%%%%%%%%%%%%%%%%%%%%%%%

\begin{center}
{\bf   Matrix oscillator and Laughlin Hall states}
 
\bigskip

S.Meljanac {\footnote{e-mail: meljanac@irb.hr}} \hspace{0.3cm} and \hspace{0.3cm}  
A. Samsarov{ \footnote{e-mail:asamsarov@irb.hr}}\\
 Rudjer Bo\v{s}kovi\'c Institute, Bijeni\v cka  c.54, HR-10002 Zagreb,
Croatia\\[3mm]

\bigskip

\end{center}
\setcounter{page}{1}
\bigskip

%%%%%%%%%%%%%%%%%%%%%%%%%%%%%%%% ABSTRACT %%%%%%%%%%%%%%%%%%%%%%%%%%%%%%%%%%%%

\begin{center}
{\bf   Abstract 

\bigskip}

\end{center}

 We propose a quantum matrix oscillator as a model that provides the
 construction of the quantum Hall states in a direct way. A connection
 of this model to the regularized matrix model introduced by
 Polychronakos is established .
By transferring the consideration to the Bargmann representation with the help of a particular similarity
 transformation, we 
 show that the quantum matrix oscillator describes the quantum mechanics of electrons
 in the lowest Landau level with the ground state described by
the Laughlin-type wave function. 
 The equivalence with the Calogero model in one dimension is emphasized. It is shown that the quantum matrix oscillator
and the finite matrix Chern-Simons model have the same spectrum on the singlet state sector.

%%%%%%%%%%%%%%%%%%%%%%%%%%%%%%%%%%%%%%%%%%%%%%%%%%%%%%%%%%%%%%%%%%%%%%%%%%%%%%%
%%%%%%%%%%%%%%%%%%%%%%%%%%%%%%%%%%%%%%%%%%%%%%%%%%%%%%%%%%%%%%%%%%%
\bigskip
PACS number(s): 03.65.Fd, 03.65.Sq, 05.30.Pr \\
\bigskip
\bigskip
Keywords: Matrix oscillator, Bargmann representation, Laughlin states .

%%%%%%%%%%%%%%%%%%%%%%%%%%%%%%%%%%%%%%%%%%%%%%%%%%%%%%%%%%%%%%%%%%%

\newpage

%%%%%%%%%%%%%%%%%%%%%%%%%%%%%%%%%%%%%%%%%%%%%%%%%%%%%%%%%%%%%%%%%%%

%%%%%%%%%%%%%%%% SECTION 1 : Introduction %%%%%%%%%%%%%%%%%%%%%%%%%%%%%%%%%%%%%

%%%%%%%%%%%%%%%%%%%%%%%%%%%%%%%%%%%%%%%%%%%%%%%%%%%%%%%%%%%%%%%%%%%

\section{Introduction}
The finding of quantum levels of nonrelativistic electrons in a uniform magnetic field is a well-known problem in
quantum mechanics and extends to studying the physics of the quantum Hall effect.
 The physics of electrons in the lowest Landau
level exhibits some interesting features, the example of which is the occurrence of the incompressible fluidlike [1]
states of condensed electrons whose excitations have fractional charge and obey fractional statistics [2,3]. These states 
appear only when the electron densities are certain rational fractions of the density corresponding to a fully filled
lowest Landau level and the gap in their excitation spectrum gives rise to the experimentally observed fractional
 quantum Hall effect. They are described
by the Laughlin wave functions [4]. The tools for studying the exactness and universality of the Laughlin wave functions
are offered in a natural way in the realm of matrix models [5].

One can argue about using noncommutative physics for describing real physical systems, such as the quantum Hall fluid.
The natural realization of noncommutative space is provided by the planar coordinates of quantum
particles moving in a constant magnetic field.
Recently, an attempt was made by Susskind [6] to describe the incompressible quantum Hall fluid in terms of the 
noncommutative Chern-Simons theory on the plane, the approach that has the connection to an analogy between the physics
of electrons in a strong magnetic field and the properties of D-branes in string theory [7].
The dynamics of quantum Hall fluids in the framework of noncommutative field theory was treated in [8,9].

As the Chern-Simons theory on the plane necessarily describes a spatially infinite quantum Hall system, it was also
of interest to find a description of finite systems with a finite number of electrons and this was achieved
by the model introduced by Polychronakos [10]. Such a regularized model, proposed as a theory of finite matrices with 
additional boundary vector fields,
has provided a description of the quantum Hall droplet and its boundary  excitations [11]. The quasiparticle
and quasihole states were explained in terms of Schur functions within an algebraic approach [12].

The finite matrix Chern-Simons model is described by two
matrices $ \; X_{1}, \; X_{2} \; $ or $ \; A, \; A^{\dagger}. $ It was shown [12] that  both these matrices
 could not be
diagonal simultaneously with some operators on the diagonal.
This would lead to inconsistencies and to only two towers of states of the Bose and Fermi type, respectively. 
 There was also a problem
with the construction of the general Laughlin states [13].
However, the strong connection of the matrix Chern-Simons model with the Calogero model and the quantum Hall effect was
 pointed out in [10-13].

Recently, a quantum matrix oscillator was proposed and its equivalence to the Calogero-type
 models was established [14,15]. The classical version of the matrix oscillator was introduced in [16]
and the path integral quantization of this model was performed in [17].

In this letter we propose a quantum matrix oscillator and establish
its connection to the finite matrix Chern-Simons model introduced
 by Polychronakos. We use the matrix oscillator model [14] to find the
 physical states of electrons in the lowest Landau level.
 The ground states are Laughlin-type states and the analysis leading to this result, together
 with the construction of the excited states, relies heavily on the consideration that is carried out in the Bargmann
 representation. The main point here is to reduce the eigenvalue problem to a much simpler one and then to transfer
 the obtained results back to the original problem, with the help of a conveniently
 constructed similarity transformation.
Although the analysis is performed for the one-dimensional case only,
it can as well be straightforwardly extended to two and higher dimensions  as long as identical particles are considered.
 As a consequence, the results obtained
can be analytically continued onto the whole complex plane incorporating in such a way the wave functions of the
true Laughlin form that depend on complex variables.
The relevance of the matrix oscillator model to the quantum Hall
physics has been emphasized throughout the procedure.

The paper is organized as follows. In the Section 2 we introduce the
matrix oscillator model and make a connection to the finite matrix
model. The next step is made in Section 3 where the
equation of motion stemming from the matrix model action is recognized as the
quantization condition imposed on the matrix coordinates of the
electrons.
After finding the representation of the matrices $ \; X_{1}  \; $ and
$ \; X_{2},  \; $ that solve the quantization condition, in Section 4
we construct the matrix operators required for building up the Fock
space of states for the matrix oscillator model. The main result and
the crucial analysis of the paper is contained in Section 5, where the
transition to a particularly convenient Bargmann representation is
made. This enables us to identify the eigenstates of the matrix
oscillator model as the wave functions of physical states describing
electrons in the lowest Landau level, including
the ground state Laughlin wave function and excitations over the
Laughlin state.

%%%%%%%%%%%%%%%%%SECTION 2 %%%%%%%%%%%%%%%%%%%%%%%%%%%%%%%%%%%%%%%%%%%%%%%%%%%%%%%%%%%%%%%%%%%%%%%%%

\section{Matrix oscillator and action}
Let us construct an action for the matrix oscillator described by $ \; N \times N \; $ matrices
$ \; X, \; {\mathcal{P}} \; $ with operator-valued matrix elements, $ \; {({X}_{ij})}^{ \dagger } = X_{ji},
\;\; {({{\mathcal{P}}}_{ij})}^{ \dagger } = {\mathcal{P}}_{ji};  \;\; i,j = 1,2,...,N. $ We take the
matrix $ \; X \; $ to be diagonal, with real elements.
 The Hamiltonian and commutation relations [14] are then $ \; (\hbar = 1) \; $
\begin{equation}
 H = R ( \frac{1}{2m} {\mathcal{P}}^{2} + \frac{1}{2}m { \omega}^{2} X^{2} ) C,
\end{equation}
\begin{equation}
 [ X, {\mathcal{P}} ] = \imath {\mathcal{V}}, \;\;\;\;\; {\mathcal{V}} = ( 1 - \nu ){\bf{1}} + \nu {\mathcal{J}},
\end{equation}
where $ \; R = (1 ......1) \; $ is a row-vector whose all components are units, and $ \; C = R^{T} \; $
is a transpose of $ R $. Also, we have $ \; RC = N \; $ and  $ \;  CR = {\mathcal{J}}, \; $ where $ {\mathcal{J}} $
is the $ \; N \times N \; $ matrix with units at all positions.
The matrix $ \;  {\mathcal{V}} \; $ is symmetric, $ \; {\mathcal{V}}^{T} = {\mathcal{V}},  \; $ 
where $ \; \nu > -\frac{1}{N} \; $ is a real parameter and $ \; m \; $ is the mass. Generally, $\; {\mathcal{V}} \; $
is a Hermitian matrix $ \; {\mathcal{V}}^{\dagger} = {\mathcal{V}},  \; $ with $ \; \nu_{ii} = 1 \; $ and 
$ \; {\nu_{ij}}^{*} = \nu_{ji}, \; \forall i,j,$ and the effective Hamiltonian contains three-body interactions [15].

In order to describe two-dimensional systems of $ \; N \; $ charged particles with charge $ \; e \; $ in a magnetic
field $ \; B, \; $ it is convenient to define the matrix $ \; X_{1} \equiv X \; $ and  a second matrix
 $ \; X_{2} \; $ expressed in terms of $ \; {\mathcal{P}} \; $ as
\begin{equation}
  X_{2} = - \frac{1}{eB} {\mathcal{P}} = - \frac{1}{m \omega} {\mathcal{P}}, \;\;\;\;\;
\end{equation}
where $ \; \omega = \frac{ e B}{m}. $ Note that the trace $ \; Tr[X_{1}, X_{2}] \; $ is equal to
 $ \; \frac{N}{\imath eB}, \; $ in accordance with the relation (2). 

The coordinates of the electrons can be globally parametrized in a fuzzy way by introducing two $ \; N \times N \; $
Hermitian matrices $ \; X_{a}; \; a = 1,2. \; $
The action leading to the quantum matrix oscillator  is then given by
the regularized finite matrix Chern -Simons model introduced by Polychronakos
$$
 S_{M} = \frac{eB}{2} \int dt Tr[ {\varepsilon}_{ab} X_{a}( {\dot{X}}_{b} - \imath[A_{0}, X_{b}]) + 2 \theta A_{0}]
$$
\begin{equation} 
         - \frac{ \omega e B N}{2 \bar{\psi}\psi} \int dt \bar{\psi} X_{a}X_{a} \psi 
 -   \int dt  \bar{\psi} (\imath \partial_{t} + A_{0} ) \psi,
\end{equation}
where $ \; e B \theta = k, \;\; A_{0} \; $ is a matrix entering into
the above action only linearly and $ \; \psi \; ( \bar{\psi} =
{\psi^{*}}^{T}) \; $ is a boundary vector field.  The action (4) is
invariant under the transformations $ \; X_{a} \rightarrow U X_{a}
U^{-1}, \;\; \psi \rightarrow U \psi, \;\; \bar{\psi} \rightarrow
\bar{\psi} U^{-1}, \;\; A_{0} \rightarrow U A_{0} U^{-1} + \imath U
\partial_{t} U^{-1}, \;\; $ where $ \; U \; $ is a unitary matrix, $
\; U \in U(N). $ The term with $ \; \omega \; $ serves as a potential
box that keeps particles near the origin and also provides a
Hamiltonian for the theory that chooses a unique ground state, while
the last term in the action can be interpreted as a boundary term.
Also, note that the minor change is made in the harmonic term in 
respect to the action of Ref.[10], namely $ \; Tr {(X_{a})}^{2} \;$ is
replaced by $ \; \bar{\psi} {(X_{a})}^{2} \psi $. But, as these two
parts yield the same spectrum when acting on the singlet sector of the
$ \; U(N) \; $ group, this replacement essentially does not make any
 difference. The only reason for replacing the $ \; Tr {(X_{a})}^{2} \;$ 
 by $ \; \bar{\psi} {(X_{a})}^{2} \psi $ is that the later gives rise 
  to the quantum  Calogero model (in the quantum Calogero model the
  inverse square potential term has  $ \; \nu(\nu + 1) \; $ as a
  prefactor, with $ \; {\nu} \;  $ being the coupling constant),
 while the former is related
 to the classical Calogero model (this has the factor $ \; {\nu}^{2} \;  $  in
 front of the inverse square potential term).
Later, we shall see that, after a 
diagonal form of one of the matrices $ \;  X_{1} \; $ or $ \; X_{1} + \imath X_{2} \; $ is assumed,
 the boundary fields transform
into the  $ \; R, C \; $ matrices, i.e. row and column matrices defined after Eq. (2).

%%%%%%%%%%%%%%%%%%%%%%%%%%% SECTION 3 %%%%%%%%%%%%%%%%%%%%%%%%%%%%%%%%%%%%%%%%%%%%%%%%%%%%%%%%%%%%5

\section{  Gauss condition and quantization}
 The variation of the action $ \; S_{M} \; $ in the field variable $ \; A_{0} \; $ gives the equation of
motion for the time component $ \; A_{0} \; $ of the gauge field. This equation has the form
\begin{equation}
  \imath e B [X_{1}, X_{2}] + k {\bf 1 } -  \psi \bar{\psi} = 0
\end{equation}
and can be interpreted as the Gauss law.
Now we recognize  the Gauss law (5) as a quantization condition imposed on the
matrices $ \; X_{1} \; $ and $ \; X_{2}, \; $ after which their matrix
elements become operators [14]. As an additional
point, we require that one of the $ \; X_{a} \; $ matrices, say $ \; X_{1}, \; $ can be diagonalized. From Eq. (5)
it follows $ \; Tr [X_{1}, X_{2}] = \frac{Tr \psi \bar{\psi} - Nk}{\imath e B} = \frac{N}{\imath eB}, $ 
in agreement with Eqs. (2),(3). This means
that $ \; Tr \psi \bar{\psi} = N(k + 1). $ 

 At this point it is important to note that 
 certain quantization constraints $ \; (k \in Z) \; $ can be imposed on the parameter $ \; k \; $ and these may be
 justified by some group theoretic arguments [18].
 So, in further considerations, $ \; k \; $ will be an integer. The obviously redundant number of degrees of freedom
is reduced to effectively $ \; 2N \; $ phase space variables with the help of the Gauss law constraint (5)
and $ \; U(N) \; $ gauge symmetry.
At the beginning we had $\; 2N^{2} \; $ degrees of freedom and  $\; 2N \; $ 
components of the boundary complex vector. After diagonalizing $ \; X_{1}, \; $ and solving the Gauss constraint,
 we are left with  $\; 2N \; $ degrees of freedom, corresponding to $\; N \; $ electrons. 

In the action (4) we have introduced a quadratic potential
 $ \; \frac{N}{2 \bar{\psi} \psi}m {\omega}^{2} \bar{\psi} {(X_{a})}^{2} \psi \; $
which, after the diagonalization of the matrix $ \; X_{1}, \; $ becomes equal to the 
quantum matrix oscillator Hamiltonian (1). More explicitly, the unitary transformation $ \; U \; $ which diagonalizes
the matrix $ \; X_{1}, \;\; U X_{1}U^{\dagger} = X'_{1} = diag(x_{1},...,x_{N}), \; $
 will change the vector $ \; \psi \;$ into $ \; \phi = U \psi \; $ and
 the matrix $ \; X_{2} \; $ into $ \; X'_{2}, \; $  so that, after solving the quantization constraint (5), 
 it can be represented [14] with the 
following operator-valued elements:
\begin{equation}
 - \imath eB {(X'_{2})}_{ij} = ( \frac{\partial}{\partial x_{i}} + 
\sum_{k \neq i} \frac{ \lambda_{ik}}{x_{i} - x_{k}})
 {\delta}_{ij} -   \frac{1 - {\delta}_{ij}} {x_{i} - x_{j} } {\phi}_{i}{\bar{\phi}}_{j},
\end{equation}
where the eigenvalues of $ \; X_{1} \; $ can be interpreted as the particle coordinates in the $ \; x_{1} \; $ direction.
The parameters $ \; \lambda_{ik},  \;\; i,k = 1,...,N, \; $ are gauge parameters and $ \; \phi \; $ is the vector
we end up with, after the vector $\; \psi \; $ is rotated by the transformation $\; U, \; \phi = U \psi.  $
In the following, we work in the gauge where all gauge parameters $ \; \lambda_{ik} \; $ are equal to zero.
The Gauss law (5) is now just a deformed quantization condition (2) that
can be rewritten in the form
\begin{equation}
 - eB[X'_{1}, X'_{2}] = \imath {\mathcal{V'}}, \;\;\;\;\; 
 {\mathcal{V'}} = - k {\bf 1 } +  \phi \bar{\phi}.  
\end{equation}
If one of the matrices $ \; X'_{1}, X'_{2} \; $ in the relation (7) is diagonal, which is the case here, then the
consistency of the solution of the commutation relation (7) necessarily requires that the matrix
 $ \; {\mathcal{V'}} \; $ should be of the form
$ \; {\mathcal{V'}} = -k {\bf 1 } + ( k + 1) {\mathcal{J}}, $ where the matrix $ \; {\mathcal{J}} \; $
 has already been defined after Eq. (2).  
Namely, a more detailed analysis shows that  
$ \; | \phi_{i} | \equiv |(U \psi )_{i} | = \sqrt{k + 1}, \; k \geq -1 \; $ and
the residual $ \; U(1)^{N} \; $ gauge freedom can be used to choose the phase factors of 
$ \; {\phi}_{i} \; $ so that $ \; {\phi}_{i} = \sqrt{k + 1}. $  
The matrix $ \; {\mathcal{V'}} \; $ is equal to the matrix $ \; {\mathcal{V}} =
                       (1 - \nu){\bf 1} + \nu {\mathcal{J}}, \; $ where 
 $ \; \nu = k + 1. $
In the classical limit $ \; \hbar \rightarrow 0  \; $ or equivalently $ \; \nu \rightarrow \infty, \; $ we have
$ \; Tr [X_{1}, X_{2}] = 0 \; $ and the diagonal elements in $ \; {\mathcal{V'}} \; $ are equal to zero.
In regard to the parameter $ \; k, \; $ the  two specially interesting cases are when
$ \; k = -1 (\nu = 0)  \; $ and $ \; k = 0 (\nu = 1). $ The former corresponds to the Bose system and 
the latter corresponds to the Fermi system.

%%%%%%%%%%%%%%%%%%%%%%%%%%%%%%%%%%%%% SECTION 4 %%%%%%%%%%%%%%%%%%%%%%%%%%%%%%%%%%%%%%%%%%%%%%

\section{ Fock space representation}
Let us introduce matrix operators 
\begin{equation}
 {\mathcal{A}}^{\pm} = \sqrt{ \frac{m \omega }{2 }} ( X'_{1} \pm \imath X'_{2})
\end{equation}
such that the following commutation relation holds:
\begin{equation}
  [{\mathcal{A}}^{-}, {\mathcal{A}}^{+}] =   -k {\bf{1}} +  \phi \bar{\phi} =
(1-\nu) {\bf 1} + \nu {\mathcal{J}}. 
\end{equation}
Owing to the fact that the fields $ \; \phi, \bar{\phi} \; $ are proportional to $ \; R, C \;$ matrices, i.e.
$ \; \phi = \sqrt{k + 1}C, \; \bar{\phi} = \sqrt{k + 1}R,  \; $ the Hamiltonian can now be written in a way as Eq. (1),
\begin{equation}
  H = \frac{  \omega}{2(k + 1)} \bar{\phi}\{ {\mathcal{A}}^{-}, {\mathcal{A}}^{+} \} \phi =
     \frac{  \omega}{2} R\{ {\mathcal{A}}^{-}, {\mathcal{A}}^{+} \} C.  
\end{equation}
The ground state is a column vector $ \; {\parallel 0 \rangle}_{ \nu } \; $ that is
 annihilated by the operator $ \; {\mathcal{A}}^{-}, \; $
\begin{equation}
  {\mathcal{A}}^{-}{\mathcal{J}} {\parallel 0 \rangle}_{ \nu } =
 {\mathcal{A}}^{-} {\parallel 0 \rangle}_{ \nu } = 0, \;\;\;\;
{\parallel 0 \rangle}_{ \nu } \sim  C \prod_{i < j}{(x_{i} - x_{j})}^{\nu} e^{-\frac{m \omega}{2} \sum_{i}{x_{i}}^{2}}
\end{equation}
and the full Fock space [14] is given by the states
\begin{equation}
 \prod_{n} ( Tr   {({{\mathcal{A}}^{+}}^{n}{\mathcal{J}}))}^{m_{n}} {\parallel 0 \rangle}_{ \nu } =
 \prod_{n} {( \sum_{i} { (a_{i}^{+})}^{n})}^{m_{n}} {| 0 \rangle}_{ \nu },
\end{equation}
where $ \; { (a_{i}^{+})}^{n} = { (R {{\mathcal{A}}^{+}}^{n})}_{i}, \;\;
  { (a_{i}^{-})}^{n} = { ( {{\mathcal{A}}^{+}}^{n} C)}_{i}, \;\;  i = 1,...,N, \; $ with
$ \; {a_{i}}^{+}, \; {a_{i}}^{-} \; $ being the one-particle creation and annihilation operators [14] for the
Hamiltonian $ \; H. $ 
The corresponding energies are
\begin{equation}
 E_{ \{ m \} } = E_{0} + \omega \sum_{n} n m_{n},
\end{equation}
where
\begin{equation}
 E_{0} = \omega ( \frac{N}{2} +  \nu \frac{N(N - 1)}{2}  ), \;\;\;\; \nu \geq 0.
\end{equation}
Note that this spectrum is the same as that following from the term 
$ \; Tr {(X_{a})}^{2}, \; $  because the corresponding Hamiltonians act as a number operator, up to the constant
$ \;  E_{0}, \; $ on the singlet part of the Fock space of states.

%%%%%%%%%%%%%%%%%%%%%%%%%%%%%%%%% SECTION 5 %%%%%%%%%%%%%%%%%%%%%%%%%%%%%%%%%%%%%%%%%%%%%%%%5

\section{Bargmann representation}
Now we analyze the structure of energy eigenstates in the Bargmann representation, as was done for the generalized
Calogero model in arbitrary dimension [19]. Starting from the matrices $ \; X_{1}, X_{2}, \; $ we define
the combinations $ \; {\mathcal{A}}^{\pm}_{B} = \sqrt{\frac{m \omega }{2}} ( X_{1} \pm \imath X_{2}),$
where the label $ \; B \;$
indicates that we are working in the Bargmann representation, the transfer to which is realized by the similarity
transformation
\begin{equation} 
  {\mathcal{A}}^{+}_{B} =  S^{-1} {\mathcal{A}}^{+} S, \;\;\;\;
 {\mathcal{A}}^{-}_{B} =  S^{-1} {\mathcal{A}}^{-} S,
\end{equation}
where $ \; S \; $ is the following operator:
\begin{equation}
   S = e^{- \omega T_{+}} e^{- \frac{1}{2 \omega} T_{-}}.
\end{equation}
The operators $ \; T_{+}, T_{-}, \; $ together with the operator $ \; T_{0}, \; $ are the generators [14]
 of the $ \; SU(1,1) \; $ algebra and are given as follows:
$$
 T_{+} = \frac{m}{2} \bar{\psi} {{X}_{1}}^{2} \psi = \frac{m}{2} R {{X'}_{1}}^{2} C,
$$
$$
 T_{-} = - \frac{m {\omega}^{2}}{2} \bar{\psi} {{X}_{2}}^{2} \psi = - \frac{m {\omega}^{2}}{2} R {{X'}_{2}}^{2} C,
$$
\begin{equation}
 T_{0} = - \frac{ \imath m \omega}{4} \bar{\psi} ( {X}_{1} {X}_{2} + {X}_{2}{X}_{1} ) \psi =
         - \frac{ \imath m \omega}{4} R ( {X'}_{1} {X'}_{2} + {X'}_{2}{X'}_{1} ) C.
\end{equation}
 The frequency $ \; \omega \; $
is assumed to be different from zero. 
Note that the same transformation connects the Hamiltonians 
$ \; H =  \frac{  \omega}{2} R\{ {\mathcal{A}}^{-}, {\mathcal{A}}^{+} \} C \; $ and
$ \; H_{B} = \frac{  \omega}{2} R\{ {\mathcal{A}}^{-}_{B}, {\mathcal{A}}^{+}_{B} \} C, \; $ together with their
corresponding ground states, namely
\begin{equation}
  H =  S H_{B} S^{-1} = 2 \omega  S T_{0} S^{-1} , \;\;\;\;\; {| 0 \rangle}_{\nu} = S {| 0 \rangle}_{\nu}^{B}
\end{equation}
and that the commutation relation satisfied by $ \; {\mathcal{A}}^{-}_{B}, {\mathcal{A}}^{+}_{B} \; $
 is still unchanged
\begin{equation}
 [{\mathcal{A}}^{-}_{B}, {\mathcal{A}}^{+}_{B}] =    -k {\bf{1}} +  \phi \bar{\phi} =
(1-\nu) {\bf 1} + \nu {\mathcal{J}}. 
\end{equation}
As the same relation (18), up to the factor $ \; 2 \omega, \; $ is also satisfied by
the operator $ \; T_{0}, \; $ which, when rewritten in an explicit form, is equal to
 $ \; \frac{1}{2} ( \sum_{i} x_{i}\frac{\partial}{\partial x_{i}} + \frac{N}{2} ), \; $
we conclude that the Hamiltonian in the Bargmann representation $ \; H_{B} \; $
 is exactly the operator $ \; 2 \omega T_{0}. $
This does not mean that the operators $ \; {\mathcal{A}}^{-}_{B}, {\mathcal{A}}^{+}_{B} \; $ can be identified
with the matrices $ \; {X'}_{1}, {X'}_{2}. $ In other words, $ \; {\mathcal{A}}^{+}_{B} \; $ is not a diagonal
matrix, but rather it satisfies the relation $ \; R {{\mathcal{A}}^{+}_{B}}^{n} C = \sum_{i}{({a_{i}}^{+})}^{n}_{B}, \; $
where $ \; {({a_{i}}^{+})}^{n}_{B} = { (R {{\mathcal{A}}^{+}_{B}}^{n})}_{i}. $ An analogous relation holds for 
$ \; {\mathcal{A}}^{-}_{B}, \; $ namely $ \; R {{\mathcal{A}}^{-}_{B}}^{n} C = \sum_{i}{({a_{i}}^{-})}^{n}_{B}, \;\;
 {({a_{i}}^{-})}^{n}_{B} = { ( {{\mathcal{A}}^{-}_{B}}^{n} C)}_{i}. $ 
These totally symmetric combinations of operators $ \; {({a_{i}}^{+})}^{n}_{B} \; $ act as  creation operators for
 the Hamiltonian $ \; H_{B} = 2 \omega T_{0}, \; $ so that the whole Fock space for $ \; H_{B} \; $
 can be constructed by applying them to the
vacuum state
\begin{equation}
 {| 0 \rangle}_{\nu}^{B} = \prod_{i < j}{(x_{i} - x_{j})}^{\nu},
\end{equation}
which is annihilated by the covariant derivative $ \; d_{i} $
\begin{equation}
d_{i} {| 0 \rangle}_{\nu}^{B} \equiv 
   (\frac{\partial}{\partial x_{i}} -
   \nu  \sum_{l \neq i}\frac{1 } {x_{i} - x_{l} })
 \prod_{j < k}{(x_{j} - x_{k})}^{\nu} = 0.
\end{equation}
The operators $ \; x_{i}, d_{i}, \; i,j = 1,...,N \; $ satisfy the commutation relations
 $ \; [d_{i}, x_{j}] = \delta_{ij}; \;\; [d_{i}, d_{j}] = 0 \; $ 
 and the Hamiltonian  $ \; H_{B} \;$ 
 can be expressed in terms of them in the following way:
\begin{equation}
 H_{B} = 
 E_{0} + \omega \sum_{i}x_{i} d_{i}.
\end{equation}
As a consequence, we have the following set of relations:
$$
  [H_{B}, x_{i}] = x_{i}, 
$$
\begin{equation}
  [H_{B}, d_{i}] = - d_{i},
\end{equation}
which allows us to interpret $ \; x_{i}, d_{i} \; $ as a pair of creation and annihilation operators for the 
Hamiltonian $ \; H_{B}. $ However, only totally symmetric combinations of these operators
have the physical meaning,
so the true Fock space for $ \; H_{B} \; $ is constructed by applying the operators
\begin{equation}\begin{array}{l}
  B_{n}^{+} \equiv \sum_{i} {x_{i}}^{n}, \\
  B_{n}^{-} \equiv \sum_{i} {d_{i}}^{n}
\end{array}\end{equation}
to the vacuum (20). In view of the arguments just stated, the sums of powers of the operators 
 $ \; {({a_{i}}^{+})}_{B}, \; {({a_{i}}^{-})}_{B} \; $ in the Bargmann representation
are in fact reduced to $ \; \sum_{i}{({a_{i}}^{+})}^{n}_{B} = \sum_{i} {x_{i}}^{n}, \; $ i.e.
$ \; \sum_{i}{({a_{i}}^{-})}^{n}_{B} = \sum_{i} {d_{i}}^{n}. $  

The $ \; SU(N) \; $ invariant ground-state vacuum in the Bargmann representation, for a fixed $ \nu, $ is 
\begin{equation}
 {{| 0  \rangle }_{\nu}}^{B} \sim   ({\varepsilon}_{i_{0}...i_{N-1}}
 \prod_{k = 0}^{N-1} { {( R {{\mathcal{A}}^{+}_{B}}^{k} )}_{i_{k}} ) }^{\nu} {| 0 \rangle}_{0}^{B} \equiv
  ({\varepsilon}_{i_{0}...i_{N-1}}
 \prod_{k = 0}^{N-1}  { { (a_{i_{k}}^{+})}^{k}_{B}) }^{\nu} {| 0 \rangle}_{0}^{B},
\end{equation}
where $ \; { (a_{i_{k}}^{+})}^{k}_{B} = { (R {{\mathcal{A}}^{+}_{B}}^{k})}_{i_{k}}, \;\;
  { (a_{i_{k}}^{-})}^{k}_{B} = { ( {{\mathcal{A}}^{+}_{B}}^{k} C)}_{i_{k}}, \;\;  i_{k} = 1,...,N, \; $ with
$ \; { ({a_{i_{k}}}^{+})}_{B}, \; {({a_{i_{k}}}^{-})}_{B} \; $ being the one-particle creation
 and annihilation operators [14] for the
Hamiltonian $ \; H_{B}. $ Owing to the fact that we know the transformation from $ \; H_{B} \; $ to $ \; H, \; $
we also know the transformation between the corresponding ladder operators
\begin{equation}
  S \sum_{i} {({a_{i}}^{\pm})}^{n}_{B} S^{-1} = \sum_{i} {(a_{i}^{\pm})}^{n}.
\end{equation}
As a consequence, in the Bargmann representation the expression for the 
ground-state takes on the form 
\begin{equation}
  {\parallel 0  \rangle }_{\nu}^{B} = S^{-1} {\parallel 0  \rangle }_{\nu} \sim C {\prod_{i < j}}
        {( x_{i} - x_{j} )}^{\nu} \equiv C {| 0 \rangle}_{\nu}^{B}, \;\;\;\; \nu \geq 0,
\end{equation}
with $ \; {\mathcal{A}}^{-}_{B} \parallel 0 { \rangle }_{\nu}^{B} = 0. $  Then all states, Eq. (12),
 in the Bargmann
representation, with the covariant matrix derivative, Eq. (21), can be represented as
\begin{equation}
 C \prod_{n} { ( \sum_{i} {x_{i}}^{n}) }^{m_{n}} \prod_{i < j} {( x_{i} - x_{j} )}^{\nu}.
\end{equation}
For example, the quasihole state in the Bargmann representation is
 $ \; \prod_{i = 1}^{N} (z - {x_{i}}) \prod_{i < j}{(x_{i} - x_{j})}^{\nu}, \; $ where $ \; z \; $
 is a complex number. 
Note that this result is the same as the one we would get if we assumed the diagonal form for the operator
 $ \; {\mathcal{A}}^{+}_{B}, \; $  i.e., $ \; {\mathcal{A}}^{+}_{B} \equiv X'_{1}, \;\;
 {\mathcal{A}}^{-}_{B} \equiv X'_{2} \; $ with $ \; x_{i} \; $ as real variables.

The lowest state in a given tower with fixed $ \nu $ is just a Laughlin wave function.
For $ \; \nu = 0 (k = -1), \; $ both operators $ \; {\mathcal{A}}^{\pm} \; $ are diagonal and
the system is equivalent to $ \; N \; $ ordinary  one-dimensional harmonic oscillators. Generally,
the Laughlin wave function exponent $ \; \nu = k + 1 \; $
is an integer number and, particularly, if $ \; \nu \; $ is
even, the system behaves as the Bose system, and for $ \; \nu \; $
odd, it behaves as the Fermi system.
Therefore, we have shown that the transition to the Bargmann
representation allows us to eliminate the gauge degrees of freedom and
reduce the finite Chern-Simons matrix model to the quantum mechanics
of $ \; N \; $ variables with ground state given by the Laughlin wave function.

 Finally, we point out that our two-dimensional system of $ \; N \; $ particles is
completely equivalent to the quantum matrix oscillator which was shown [14] to be completely equivalent to the
one-dimensional Calogero $ N $ -body system of identical particles.
Hence, the above polynomials, Eq. (28), can be written for the Calogero system 
and the corresponding ground states are of the Laughlin type,
 $ \; \prod_{i < j}{(x_{i} - x_{j})}^{\nu}, \;\; \nu > - \frac{1}{N}, \;  $
 with the covariant derivative of the form
\begin{equation}
  d_{i} =
   \frac{\partial}{\partial x_{i}} -
   \nu  \sum_{l \neq i}\frac{1 } {x_{i} - x_{l} }, \;\;\;\;\;\;\;
d_{i} {| 0 \rangle}_{\nu}^{B} \equiv d_{i} \prod_{j < k}{(x_{j} - x_{k})}^{\nu} = 0.
\end{equation}
As we have seen, the operators $ \; x_{i}, d_{i}, \;\; i,j = 1,...N,  \; $ represent a pair
 of creation and annihilation operators for the Hamiltonian  $ \; H_{B} \;$ which
 can be expressed in terms of them as
\begin{equation}
 H_{B} = 
 E_{0} + \omega \sum_{i}x_{i} d_{i}.
\end{equation}
Note that for $ \; \nu < 0 \; $ and $ \; \nu > - \frac{1}{N}, \; $ the wave functions diverge for coincident points,
 but are still quadratically integrable [15,19].

%%%%%%%%%%%%%%%%%%%%%%%%%%%%%%% SECTION 6 %%%%%%%%%%%%%%%%%%%%%%%%%%%%%%%%%%%%%%%%%

\section{Conclusion}
We have shown that the quantum matrix oscillator introduced in [14] leads to Laughlin ground states in the 
Bargmann representation. This has provided us with the possibility of reducing the problem stemming from the action (4)
 to a simpler one which has made the underlying stucture more obvious and has provided 
an immediate physical interpretation of the results obtained.
Due to the fact that the above procedure can be straightforwardly extended to higher dimensions, the results obtained
can be analytically continued onto the whole complex plane, so as to incorporate the wave functions that depend on the
 complex variables.
In our approach the Gauss condition is interpreted as the deformed matrix 
quantization condition, and the Laughlin wave function exponent $ \; \nu = k + 1 \; $ is an 
integer number. The additional restriction on the Laughlin exponent $ \; \nu > -\frac{1}{N} \; $ 
is dictated by the existence of the ground state and, because $ \; \nu \; $ is an integer,
 it reduces to the relation $ \; \nu \geq 0. $
In contrast to the model of Ref. [10], where the term  $ \; Tr {(X_{a})}^{2} \; $ appears, here we have the  
Hamiltonian with the
 $ \; \frac{N}{\bar{\psi} \psi} \bar{\psi} {(X_{a})}^{2} \psi \; $
 term, but this does not introduce any difference because both terms
 give the same spectrum when acting on the singlet sector of the $ \; U(N) \; $ group.
In both cases there is no way to incorporate the Jain states [20] as long as $ \; k \; $ is an integer.
There is a complete equivalence to the one-dimensional Calogero $ N - $ body system [14,15], with the ground state of 
the Laughlin type.
Therefore, the Bargmann space analysis allowed us to reduce the
regularized finite Chern-Simons matrix model to the $ \; 1 - \; $
dimensional quantum $ \; N - \; $ body problem with ground state given
by the Laughlin wave function.
 We hope that our simple  
 quantum matrix oscillator may be relevant to the application
to the quantum Hall physics, particularly if extended and applied to higher dimensions [21,22].
Similar results were obtained in Ref. [23] by using the path integral approach and $ \; \em{W_{\infty}} \; $
symmetry analysis.

%%%%%%%%%%%%%%%%%%%%%%%%%%%%%%%%%%%%%%%%%%%%%%%%%%%%%%%%%%%%%%%%%%%%%%%%%%%%%%%%%%%%%%%%%%%%%%%%%%%

%%%%%%%%%%%%%%%%%%%%%%%%%%%%%%%%%%%%% Acknowledgment   %%%%%%%%%%%%%%%%%%%%%%%%%%%%%%%%%%%%%%%%%%%%%%%

%%%%%%%%%%%%%%%%%%%%%%%%%%%%%%%%%%%%%%%%%%%%%%%%%%%%%%%%%%%%%%%%%%%%%%%%%%%%%%%
{\bf Acknowledgment}\\
This work was supported by the Ministry of Science and Technology of the Republic of Croatia under 
contract No. 0098003.

\newpage

%%%%%%%%%%%%%%%%%%%%%%%%%%%%%%%%%%%%%%%%%%%%%%%%%%%%%%%%%%%%%%%%%%%%%%%%%%%%%%%%%%%%%%%%%%%%%%%%%%%

%%%%%%%%%%%%%%%%%%%%%%%%%%%%%%%%%%%%%  REFERENCES  %%%%%%%%%%%%%%%%%%%%%%%%%%%%%%%%%%%%%%%%%%%%%%%

%%%%%%%%%%%%%%%%%%%%%%%%%%%%%%%%%%%%%%%%%%%%%%%%%%%%%%%%%%%%%%%%%%%%%%%%%%%%%%%

\end{document}